\documentclass[aps,prb,twocolumn,amsmath,amssymb,showpacs,prb,superscriptaddress]{revtex4-1}

\usepackage{graphicx}
\usepackage{color}

\makeatletter

\@ifundefined{textcolor}{}
{
 \definecolor{BLACK}{gray}{0}
 \definecolor{WHITE}{gray}{1}
 \definecolor{RED}{rgb}{1,0,0}
 \definecolor{GREEN}{rgb}{0,1,0}
 \definecolor{BLUE}{rgb}{0,0,1}
 \definecolor{CYAN}{cmyk}{1,0,0,0}
 \definecolor{MAGENTA}{cmyk}{0,1,0,0}
 \definecolor{YELLOW}{cmyk}{0,0,1,0}
}

\begin{document}

%
\title{Momentum dependence of the superconducting gap and in-gap states in MgB$_2$ multi-band superconductor}
%
%
%
\author{Daixiang Mou}
\author{Rui Jiang}
\author{Valentin Taufour}
\author{S. L. Bud'ko}
\author{P. C. Canfield}
\author{Adam Kaminski}
\affiliation{Division of Materials Science and Engineering, Ames Laboratory, U.S. DOE, Ames, Iowa 50011, USA}
\affiliation{Department of Physics and Astronomy, Iowa State University, Ames, Iowa 50011, USA}

%

\begin{abstract}
We use tunable laser based  Angle Resolved Photoemission Spectroscopy to study the electronic structure of the multi-band superconductor, MgB$_2$. These results form the base line for detailed studies of superconductivity in multi-band systems. We find that the magnitude of the superconducting gap on both $\sigma$ bands follows a BCS-like variation with temperature  with $\Delta_0$ $\sim$ 7 meV. The value of the gap  is  isotropic within experimental uncertainty and in agreement with pure a s-wave pairing symmetry.  We also observe in-gap states confined to k$_F$ of the $\sigma$ band  that occur at some locations of the sample surface. The energy of this excitation, $\sim$ 3 meV, is inconsistent with scattering from the $\pi$ band.
\end{abstract}
\maketitle

Multi-band superconductors have attracted renewed interest because of the recently discovered iron based high temperature superconductors, whose pairing symmetry and mechanism  are subjects of a lively debate\cite{Johnston2010,Stewart2011}. Superconductivity in multi-band systems was already considered more than half a century ago, immediately after formulation of the BCS theory, in light of superconductivity of transition metals\cite{Suhl1959}. It became apparent early on that in such a system, the magnitude of the order parameter may vary for different bands. On the experimental side, beside iron based superconductors, several compounds are clearly identified as multi-band superconductors, including NbSe$_2$\cite{Kohsaka09032007}, YNi$_2$B$_2$C\cite{Cava1994}, A$_3$C$_{60}$\cite{Hebard1991} and, perhaps most clearly and cleanly, MgB$_2$\cite{Nagamatsu2001}. MgB$_2$ is one of the most  studied multi-band superconductor because of its relatively high transition temperature (T$_C$ $\sim$ 40K) and promise as an applied superconductor\cite{Canfield2003}. The superconducting properties of MgB$_2$ agree well with  BCS/Eliashberg theory\cite{Budko2001,Hinks2001,Kortus2001,An2001,Liu2001,Choi2002}, and it provides an ideal playground for both theoretical and experimental studies. Since its discovery, MgB$_2$ become regarded as ``the'' prototypical multi-band BCS superconductor.

The crystal structure of MgB$_2$ is very simple, with alternating graphite-like boron and magnesium layers\cite{Nagamatsu2001}. Four electronic bands that cross the Fermi level (E$_F$) can be categorized into two groups: two quasi 2D $\sigma$ bands around $\Gamma$ and two 3D $\pi$ bands near Brillouin zone boundary \cite{Kortus2001}. From theory point of view,  MgB$_2$ is usually described as two band ($\sigma$ and $\pi$) superconductor. The pairing is caused by coupling of the electron to E$_{2g}$ phonon mode. However the coupling strength  is quite different for the two types of bands, giving rise to different magnitudes of the superconducting gaps\cite{Kortus2001,An2001,Liu2001,Choi2002}. The presence of two different values of the superconducting gap was confirmed by several spectroscopic measurements\cite{Giubileo2001,avarone2002,Chen2001,Tsuda2001}. Previous Angle Resolved Photoemission Spectroscopy (ARPES) measurements have revealed that the gap size on $\sigma$ band is $\sim$6 meV and the one on $\pi$ band is $\sim$3 meV\cite{Uchiyama2002,Souma2003,Tsuda2003}, which is roughly consistent with theoretical calculations. However these experiments only measured the magnitude of the supercondcuting gap at a single K point on each Fermi surface (FS) sheet. No information on how the gap varies along one FS has been reported. Electronic Raman measurements with two polarization suggests that the gap anisotropy is only 0.4 - 0.6 meV along $\sigma$ FS\cite{Quilty2002}.
Such information is important to understand role of impurities and interband scattering in prototypical multiband  superconductor.


In this paper, we use ultra high resolution tunable laser based ARPES\cite{Jiang2014} to study superconducting gap properties of MgB$_2$. Due to limits on the accessible parts of the Brillouin zone, the $\pi$ band cannot be measured using low photon energies. We concentrate our measurements on two $\sigma$ bands around $\Gamma$ point.  Both the temperature and momentum dependence of the gap structure are systematically studied. Our results show that the gap size on both $\sigma$ FS sheets are nearly isotropic, consistent with theoretical predictions and directly illustrate s-wave pairing symmetry in MgB$_2$. Surprisingly, we also discovered electronic excitation inside superconducting gap below T$_C$ at some locations on the sample surface.

\begin{figure*}[htbp]
\centering
\includegraphics[width=0.9\textwidth]{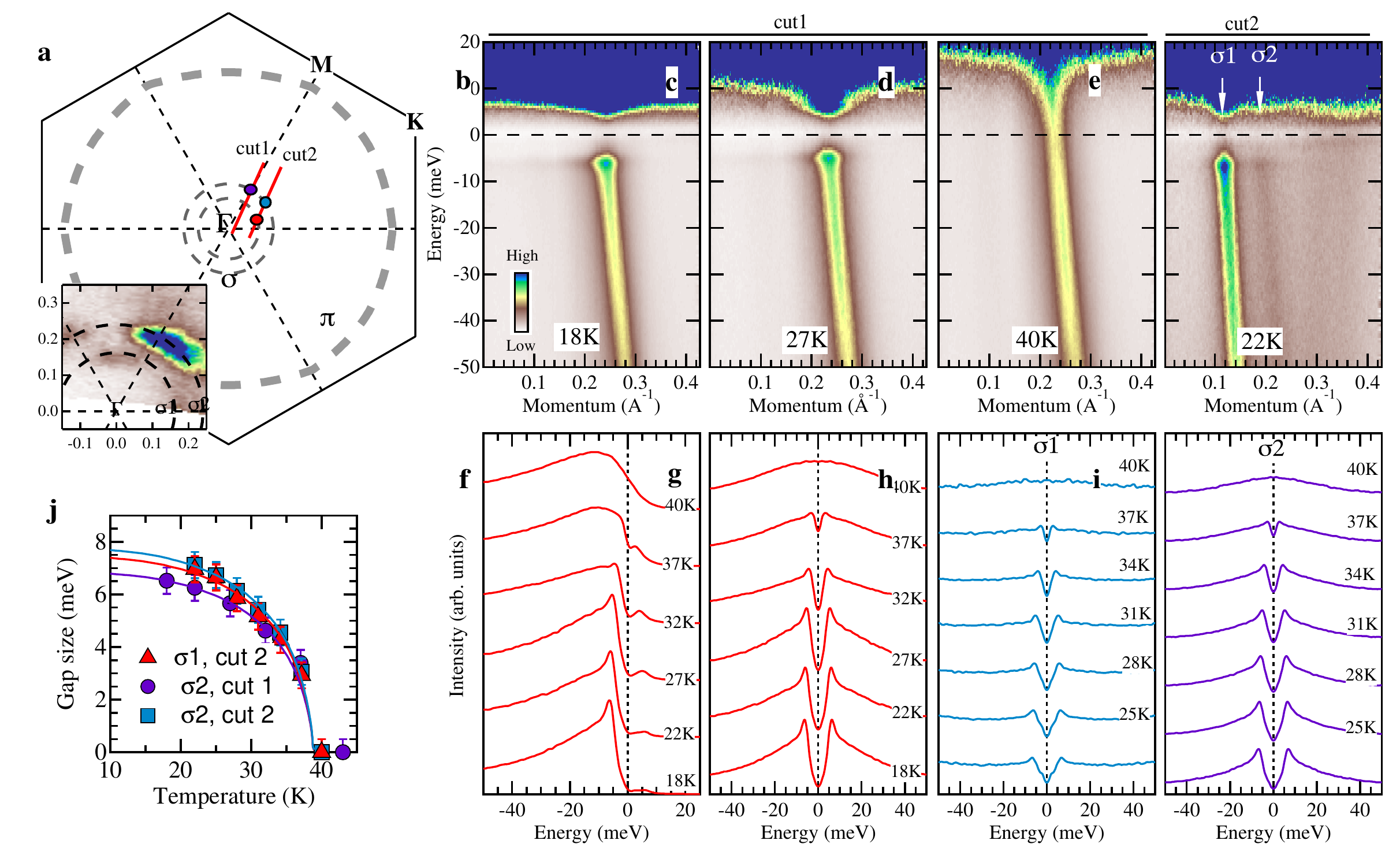}
\caption{(Color online) Temperature dependence of the superconducting gap in MgB$_2$. (a) Sketched FS topology from calculation. Insert shows the measured FS intensity map close to Brillouin zone center. (b)-(d) ARPES intensity divided by the Fermi function along cut1 ($\Gamma$-M direction) at 18K, 27K and 40K respectively. (e) ARPES intensity divided by the Fermi function along cut2 measured at 22K. Locations of cuts are illustrated in (a).  (f) Temperature dependence of EDC's at k$_F$ of cut1. (g) Symmetrized EDCs from panel (f). (h)-(i) Symmetrized EDCs at k$_F$ of cut2 for $\sigma1$ and $\sigma2$ bands respectively (j)  Temperature dependence of the superconducting gap extracted from (g)-(f) color/shape coding shown in panel (a). Solid lines indicate BCS prediction with corresponding $\Delta_0$.}
\end{figure*}

 MgB$_2$ single crystals with T$_C$ = 39 K were grown by a high pressure synthesis technique similar to that described in Ref. 23. The typical size of the samples used in our measurements is $\sim$$0.5\times0.5\times0.3$ mm$^3$. Samples were cleaved \emph{in situ} at base pressure lower than 8 $\times$ 10$^{-11}$ Torr. ARPES measurements were carried out using a laboratory-based system consisting of a Scienta R8000 electron analyzer and tunable VUV laser light source\cite{Jiang2014}. All data were acquired using photon energy of 6.7 eV, corresponding to $K_z=0.22 \pi/c$ (inner potential v$_0$ $\sim$13 eV is estimated from ref.\cite{petaccia2006characterization}). The energy resolution of the analyzer was set at 1 meV and angular resolution was 0.13$^\circ$ and $\sim$ 0.5$^\circ$ along and perpendicular to the direction of the analyzer slits, respectively. Samples were cooled using a closed cycle He-refrigerator and  the sample temperature was measured using a silicon-diode sensor mounted on the sample holder. The energy corresponding to the chemical potential was determined from the Fermi edge of a polycrystalline Au reference in electrical contact with the sample. The absence of aging effects was verified by thermal cycling. The consistency of the data was confirmed by measuring several samples.

The diagram of the Fermi surface and intensity plot at E$_f$ are shown in Fig. 1a. The ARPES intensity divided by Fermi function plots for several temperatures and two cuts are shown in Figs. 1b-e. In the  data along cut1 (Figs. 1b-d), there is only one hole-like band visible that corresponds to $\sigma2$ crossing. Intensity of the inner $\sigma1$ band is almost completely suppressed due to matrix elements. Along cut2 however (Fig. 1e), both $\sigma$ bands are visible, and the the intensity of the inner $\sigma1$ band is much stronger than the $\sigma2$. From the location of the Fermi crossing point (k$_F$), we calculate that the radii of two $\sigma$ FS sheets are $\sim$ 0.2 {\AA}$^{-1}$ and $\sim$ 0.25 {\AA}$^{-1}$ respectively.  If we ignore the small warping of these two sheets along K$_z$, we can estimate carrier concentration at 0.069 hole/cell for inner FS sheet and 0.108 hole/cell for outer one, consistent with previous quantum oscillation results\cite{Yelland2002,Carrington2003}.

We next turn to investigating the magnitude of the superconducting order parameter on the two $\sigma$ bands. The data in Fig. 1d was measured above  T$_c$ and was divided by the Fermi function. It shows a single band nicely crossing the E$_f$ characteristic of a metallic state. The data in Figs. 1b, c and e were acquired below  T$_c$ and shows the absence of intensity close to E$_f$ and increased intensity at  energy of $\sim\pm$7 meV which is the hallmark signature of the opening of a superconducting gap. The intensity at 7 meV above the E$_f$ is enhanced due to thermal excitations above 2$\Delta$, which is clearly visible especially at higher temperatures (Fig. 1c). The EDCs at the outer $\sigma$ band are shown in Fig. 1f. When the sample is cooled below T$_c$, a sharp coherent peak emerges at the energy  equal to the value of the superconducting gap and intensity close to E$_f$ is suppressed. Above E$_F$ a  smaller peak is observed due to thermal excitations above 2$\Delta$. These features are consistent with formation of Bogoliubov quasiparticles\cite{Matsui2003}. The observation of the upper branch of Bogoliubov quasiparticle band is very useful to double check the location of the chemical potential which should be centered between the two peaks as shown in Fig. 1f. In order to quantify the value of the superconducting gap we symmetrize EDCs at k$_F$ (Fig. 1g-i) and fit them with the minimal BCS model\cite{Norman1998}. Using this procedure we determine the gap magnitude at three k$_F$ points on the two FS sheets. Positions of each points in Brillouin Zone are marked in Fig. 1a. The resulting gap sizes for each temperature are shown in Fig. 1j.
All three superconducting gap magnitudes follow BCS-like temperature dependences very well with $\Delta_0$ $\sim$ 7$\pm$0.5 meV, which is slightly larger than previous ARPES results\cite{Souma2003,Tsuda2003}, but more consistent with STM results\cite{Iavarone2002} and theory calculation\cite{Choi2002}.
The gap-to-temperature ratio 2$\Delta_0$/K$_B$T$_C$ is 4.2, slightly larger than the standard single-band weak coupling BCS value 3.5, which is likely due to presence of multiple bands.

\begin{figure}[htbp]
\centering
\includegraphics[width=\columnwidth]{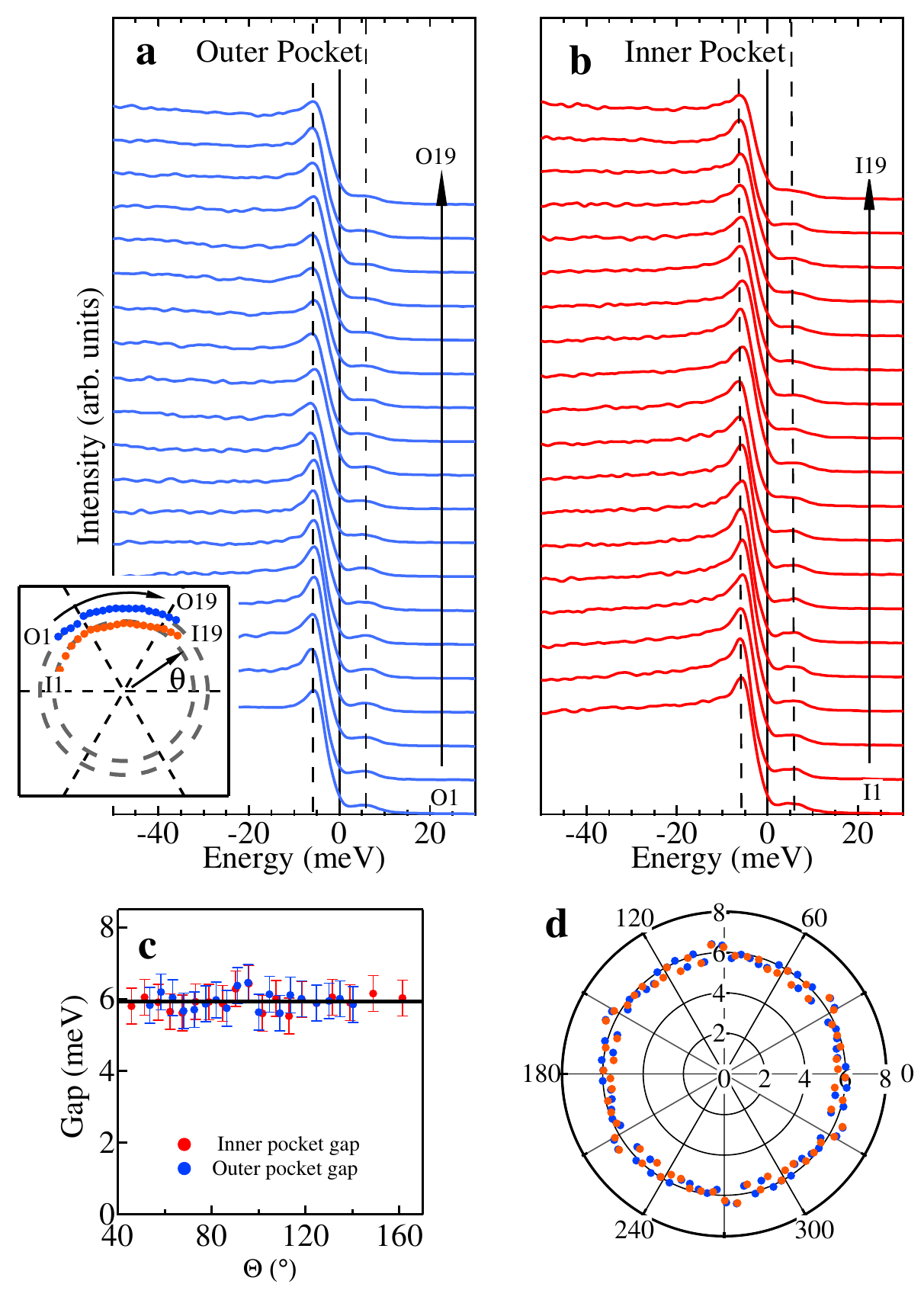}
\caption{(Color online) Momentum dependence of the superconducting gap in MgB$_2$. (a)-(b) EDCs at k$_F$ along two $\sigma$ FS sheets. The 19 momentum locations in Brillouin Zone are marked in left-bottom insert. Data were measured at 25K. (c) Angle dependence of the superconduting gap. (d) Same data as (c) plotted in polar coordinates after being six-fold symmetrized. }
\end{figure}

Previous ARPES measurements revealed that the $\sigma$ and $\pi$ FS sheets have different magnitudes of the order parameter\cite{Souma2003}. But it is not known if and how this quantity varies around the respective Fermi surface sheets. To investigate this we measure the EDCs at k$_F$  around each of the $\sigma$ sheets and plot them in Fig. 2a, b. We also extract the value of the SC gap following the same procedure as explained above. Measurements were performed at 25K to take advantage of both upper and lower Bogoliubov quasiparticle branches. The resulting values of the superconducting gap are shown in Fig. 2c, d. The gap magnitudes along both $\sigma$ FS are nearly isotropic and equal to $\sim$6 meV (for T=25K). We did not observe any systematic changes in this value within our $\pm$0.5 meV error bars. This result provides direct evidence that MgB$_2$ is a pure s-wave superconductor. Detailed gap calculation shows the average gap size on the inner pocket should be about 0.5 meV larger than that on the outer pocket\cite{Choi2002}. However at present our experimental accuracy is not sufficient to verify this prediction. Also, since both $\sigma$ bands have the same orbital character, the gap difference may be possibly  suppressed by scattering between these two FS sheets\cite{Mazin2002}.

\begin{figure}[htbp]
\centering
\includegraphics[width=\columnwidth]{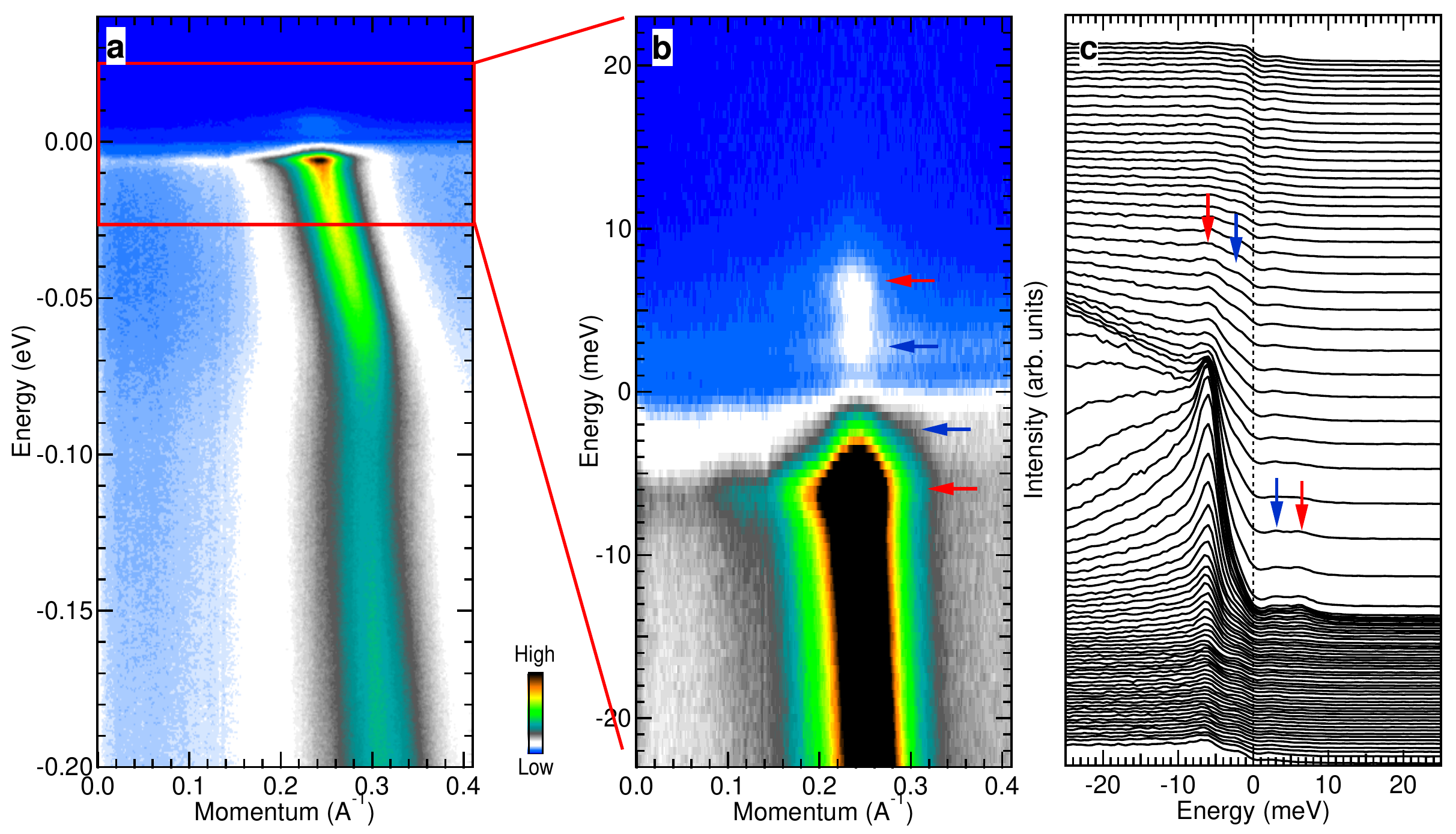}
\caption{(Color online) Inside gap state in superconducting state. (a) Measured band structure of cut1( as illustrated in fig. 1a) at 18K. (b) Expanded area close to E$_F$. Color scale is adjusted to highlight structure above E$_F$. (c)  EDCs corresponding to data in (b). Bogoliubov quasiparticle peaks and inside gap structures are marked with red and blue arrows respectively.}
\end{figure}

\begin{figure*}[htbp]
\centering
\includegraphics[width=6in]{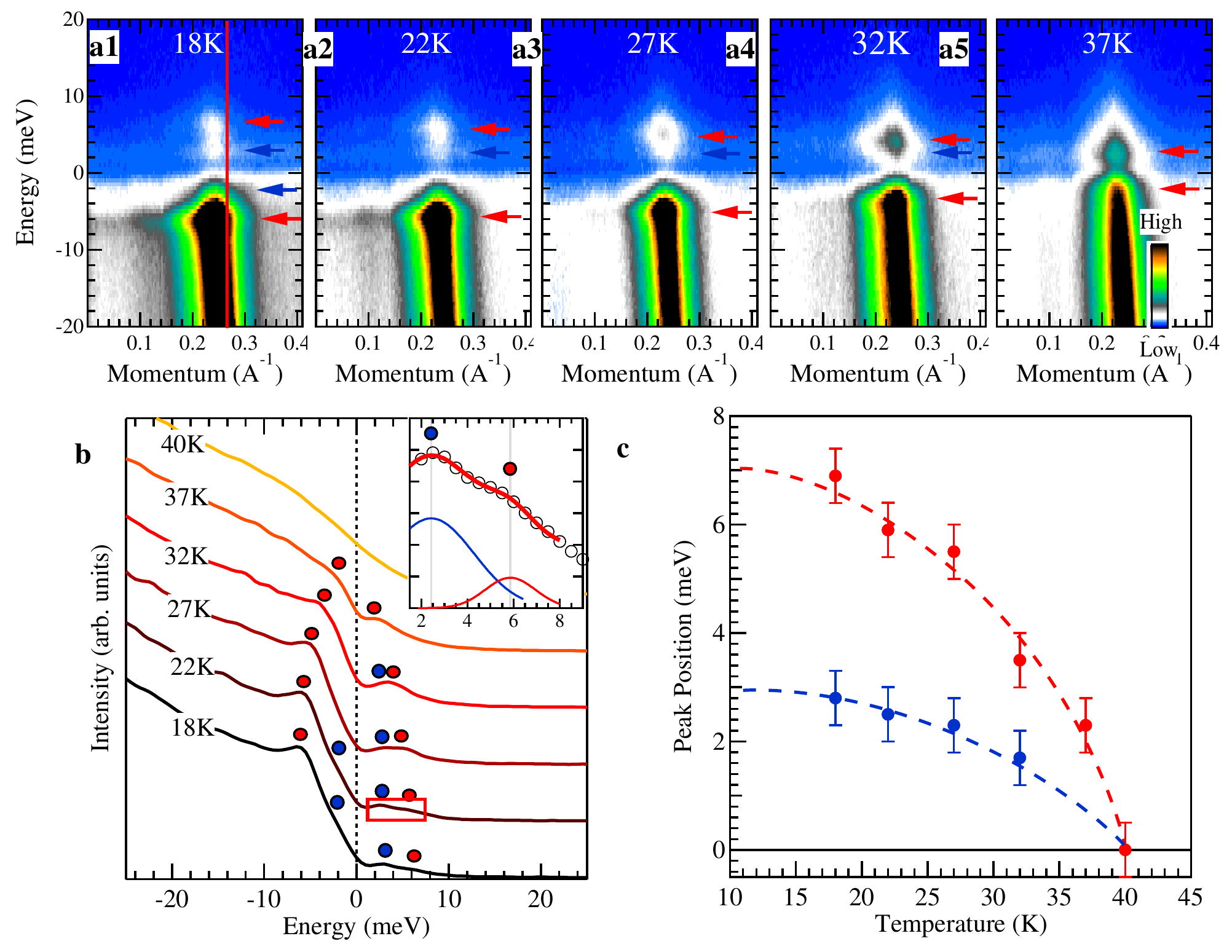}
\caption{(Color online) Temperature dependence of the  in-gap state. (a1)-(a5) ARPES intensity measured at cut 1 for various temperatures. (b)  EDCs slightly off k$_F$ (location marked with red line in a1) for various temperatures along line marked in (a1). Inset shows fitting of the EDC above E$_F$ with two gaussian peaks to extract value of the two gaps. (c) Extracted temperature dependence of the peak positions of Bogoliubov quasiparticle (red dots) and inside gap structures (blue dots). Two dashed lines are guide to the eye. }
\end{figure*}

In fig. 3a, we plot the measured electronic structure of cut1 in fig.1 without removing Fermi-Dirac function. A kink structure is clearly revealed at $\sim$ 70meV which is caused by electron coupling to E$_{2g}$ phonon mode. Its properties were discussed in great detail previously\cite{Mou2015}. In order to reveal the fine structure near E$_F$, we expand the image and adjust the color scale  in fig. 3b. Surprisingly, in addition to two branches of  Bogoliubov quasiparticle at $\pm$ 7 meV (marked with red arrows), two additional intensity peaks exists at $\pm$ 3 meV (marked with blue arrows). These are states that exist within the larger superconducting gap. (Here, the peak positions are determined by local maximum intensity. The same method is also used in Fig. 4b.) In order to get more information about this in-gap structure, we show  the temperature dependent data in fig.4. The color scale in fig. 4 is expanded to reveal this very weak structure. As temperature is increased, there are several noticeable changes. The most obvious one is that the position of two Bogoliubov quasiparticle branches moves closer to E$_F$ and  the intensity of the  Bogoliubov quasiparticle branch for positive binding energy is increasing. This is because the gap is closing and more thermal excitation occur at high temperature. At the same time, the position of the inside gap structure is also approaching E$_F$. But its intensity becomes weaker with increasing temperature and becomes almost invisible around T$_C$. In fig. 4b, we plot the EDCs slightly off k$_F$ (location marked with red line in Fig. 4a1) at different temperatures to quantify this interesting behavior. We move slightly off k$_F$  because the strong intensity of the main Bogoliubov quasiparticle peak would bury the weak inside gap state. Extracted temperature dependent positions of the Bogoliubov quasiparticle peak and inside gap structure are show in fig. 4c. Both of the peak positions roughly follow a BCS-like temperature dependence. The disappearance of this inside gap structure indicates it is closely related to superconductivity. Before discussing the possible origin of this new structure, let us summarize its main properties. This structure consists of two flat bands at $\omega_0 \sim \pm$ 3 meV at low temperature, giving the ratio of $\omega_0/\Delta_0 \sim 0.43$. Its intensity is mostly confined at k$_F$ in momentum space and its binding energy roughly follows BCS-like temperature dependence. We emphasize that this in-gap structure is only observed at some locations at the sample surface. So far we have not found any correlation between this in gap state and optically observed imperfections of the surface.

The presence of this structure is unlikely a trivial effect of momentum mixing (e. g. due to scattering of the photoelectrons) because it occurs only in the proximity of k$_F$. There are several possible explanations for presence of the inside gap structure in MgB$_2$. The most obvious one is inter-band scattering from the $\pi$ band due to disorder or impurities. The only problem is that previously measured value of the superconducting gap at $\pi$ band is  smaller by a factor of two i. e. $\sim$1.5 meV\cite{Souma2003}. A second possibility is that this structure is an impurity bound state in the superconducting gap. Early theoretical studies of the impurity effect on superconductivity indicate that, at proper scattering conditions between conduction electron and impurity,  some bound state will appear inside superconducting gap\cite{LUH1965,Shiba01091968,Balatsky2006}. The properties of this bound state have been intensively investigated both theoretically and experimentally because they could give some essential information of superconductivity, such as pairing symmetry. Bound states near impurities have been observed both in conventional\cite{Yazdani21031997} and unconventional superconductors\cite{Pan2000,Yang2013} in STM measurements. For a conventional superconductor like MgB$_2$ discussed here, a nonmagnetic impurity could not affect superconductivity much as predicted by early Anderson's theorem\cite{Anderson1959}, which means no bound state would be introduced by a nonmagnetic impurity. On the other hand, magnetic impurity can give rise to pairs of bound states both above and below E$_F$, also called Shiba states\cite{Shiba01091968,Shiba01071973,Moca2008}. The energy positions of the Shiba states vary with scattering details. Up to now, no such states have been reported in doped MgB$_2$.  The Shiba state would be buried under the Bogoliubov quasiparticle peak in spectroscopy measurements without momentum resolution. Electronic properties of the impurity bound state are studied in momentum integrated density of state in most of the existing literatures. Information on intensity distribution in momentum space is scarce. Whereas this scenario seems consistent with our results, one of the key unsolved problems is what plays the role of magnetic impurity. Both magnesium and boron atoms are nonmagnetic and no other magnetic element is involved in sample growth procedure\cite{Karpinski2007}. Further experimental results and theoretical calculations are needed to identify the scattering potential that plays the role of magnetic impurity.

In summary, the temperature and momentum dependent superconducting gaps, on the two $\sigma$ bands, were systematically studied. The gap size follows a BCS-like temperature dependence with $\Delta_0$ $\sim$ 7meV. The momentum dependent gap structure, on both FS sheets, is isotropic, giving direct evidence of s-wave pairing symmetry in MgB$_2$. We observed a flat-band-like inside gap structure below T$_c$. Its intensity is mostly confined close to k$_F$ in momentum space. The energy position of this structure also roughly follows a BCS-like temperature dependence. We proposed two possibilities to explain this new in-gap structure. Further experimental and theoretical investigations are needed arrive at a definitive answer. The electronic properties of MgB$_2$ revealed here form a basis for understanding of multi band superconductivity in conventional and unconventional superconductors.

Research was supported by the US Department of Energy, Office of Basic Energy Sciences, Division of Materials Sciences and Engineering. Ames Laboratory is operated for the US Department of Energy by the Iowa State University under Contract No. DE-AC02-07CH11358.

$^{*}$Corresponding author
kaminski@ameslab.gov

\bibliography{MgB2gap}

\end{document}